\documentclass[twocolumn,showpacs,preprintnumbers]{revtex4}
\usepackage{amssymb}
\usepackage{amsfonts}
\usepackage{amsmath}
\usepackage{graphicx}
\usepackage{dcolumn}
\usepackage{bm}

\input{tcilatex}

\begin{document}

\title{Quantum Look at two Common Logics: the Logic of Primitive Thinking
and the Logic of Everyday Human Reasoning}
\author{E. D. Vol}
\email{vol@ilt.kharkov.ua}
\affiliation{B. Verkin Institute for Low Temperature Physics and Engineering of the
National Academy of Sciences of Ukraine 47, Lenin Ave., Kharkov 61103,
Ukraine.}
\date{\today }

\begin{abstract}
Based on ideas of quantum theory of open systems and psychological dual
system theory we propose two novel versions of Non-Boolean logic. The first
version can be interpreted in our opinion as simplified description of
primitive (mythological) thinking and the second one as the toy model of
everyday human reasoning in which aside from logical deduction, heuristic
elements and beliefs also play the considerable role. Several arguments in
favor of the interpretations proposed are adduced and discussed in the paper
as well.
\end{abstract}

\pacs{05.40.-a}
\maketitle

\section{Introduction}

It is well known that the Boolean propositional logic which is consistent
mathematical presentation of the classical Aristotelian logic has
applications of two kinds. On the one hand it correctly takes into account
deductive elements of the everyday human reasoning and on the other hand
this logic can be considered as the relevant framework of common scientific
language both in exact sciences and humanities. For example referring to the
classical physics it may be argued that the result of any experiment
realized in this area can be described by the Boolean logic in a consistent
manner. However in the situations when distinct features of the objects
under study strongly correlate between themselves and moreover may be
incompatible the application of standard Boolean logic may lead to errors.
Just such case has place for example in quantum mechanics. Actually, as
firstly G. Birkhoff and J. von Neumann revealed \cite{1s} in quantum
mechanics, in view of existence of certain non-commuting observables
relating to the same particle such as coordinate and momentum or the
different components of the spin, the laws of the Boolean logic (in
particular the distributive law) have been failed and their generalization
i.e. quantum logic is desirable. The similar situation may arise also in
everyday life when various reasons, emotions and beliefs governing the
behavior of concrete person begin to contradict each other. In this
connection it should be noted that still in the early days of quantum theory
N. Bohr, when he formulated the Complementarity Principle, (which just
maintains the presence in physics certain supplementary properties of the
microscopic objects) underlined more than once essential analogy existing
between atomic processes and such mental phenomena as thoughts, sentiments
and acts of decision making. It is remarkable that modern cognitive
psychology using the data of a large number of experiments came to the
similar conclusions as well. Really according to the psychological dual
system theory all basic cognitive processes such as attention, memory,
learning and so on are connected with certain dual systems and dual
mechanisms. In the present paper we will interested in only human reasoning
mechanisms where two primary dual systems of interest can be specified. The
first of them (we will term it below as deductive reasoning system) is
rational, sequential and consistent but acts relatively slow and its
resources are limited by the capacity of human working memory. The second
one (we will term it as heuristic reasoning system ) is intuitive, rapid and
automatic but its activity may be biased to a large degree by emotions and
old unconscious ideas. In addition a hidden interaction exists between these
two cognitive systems which as a rule is not aware by the reasoning person.
The main goal of present paper is to demonstrate how by using some ideas,
concepts and technical tools of quantum theory of open systems (QTOS) one
can describe the dual nature of human reasoning in the framework of
consistent logical theory that generalizes and modifies the rules of
ordinary Boolean logic.

The paper is organized as follows. In Section.2 following the previous
author paper \cite{2s} we briefly outline main ideas of the approach
proposed using the simple and instructive example of probabilistic Boolean
logic (PBL) that describes the situations when in the absence of complete
information about the surrounding objects and events all human judgments
acquire inevitably probabilistic nature. To take into account this crucial
point it is convenient, instead of usual Boolean functions of discrete
variables which take only two values unit and zero, to associate with every
plausible proposition (PP) certain diagonal representative density matrix
(RDM) of some auxiliary two state quantum system, whose elements define the
plausibility of corresponding proposition. The essential and novel element
of the approach proposed is the universal and constructive algorithm (based
on QTOS) which allows one to define all logical connectives between
plausible propositions using the powerful and effective method of positive
valued (PV) quantum operations. It is worth noting that this method will be
used continually throughout the paper for the construction of various kinds
of logics. Also we study here another important problem, namely : how the
possible logical correlations between various PP may be took into account in
the approach proposed. Further in the Section.3 in order to describe the
dual nature of human reasoning we have extended our approach to the
arbitrary $2\times 2$ non-diagonal density matrices that will represent in
this case the generalized propositions (GP). We assume that non-diagonal
elements of the RDMs of such propositions relate to heuristic ( believable)
elements of human reasoning and therefore by means of them one can define
the believability of the corresponding propositions. The main problem of the
approach proposed is: whether one may specify the set of logical connectives
for similar generalized propositions and if the answer is yes in what way.
To answer this question we will use again the approved method of PV quantum
operations. It turns out however that in general case (beyond the usual PBL)
it is necessary to impose certain restrictions either on the form of
admissible quantum operations or on the form of input GP. In this connection
we proposed two possible alternatives to construct the consistent
Non-Boolean (NB) logic operating with such GP. The NB logic of the first
kind aside from negation includes only the single pair of two place
connectives which are similar to the pair of relations:
equivalence-non-equivalence in the standard Boolean logic. At first sight
such logic looks much poorer than standard Boolean logic and we will term it
as 'atomic'\ logic or prime logic. It turns out however that, starting from
GP and using only these two connectives, one may construct any prescribed
diagonal PP and after that to handle with these PP according to the known
rules of PBL .Thus PBL appears now as secondary or 'molecular'\ logic with
respect to prime or 'atomic' NB\ logicof the first kind.

We argue that described version of Non-Boolean logic of the first kind
(prime logic) can be considered as simplified version of the logic of
primitive (or mythological) thinking. In favor of such interpretation we
adduce several cogent in our opinion arguments. Finally in the Section.4. we
study the NB logic of the second kind that assumes the existence of special
primordial correlations between deductive and heuristic components of
generalized propositions .In this case using the formalism of PV quantum
operations one can specify all the same logical connectives between GP
including implication as in usual PBL .It turns out that a number of known
from everyday life curious psychological phenomena such as belief-bias
effect may be explained in the framework of this logic by natural way. Now
let us go to the concrete presentation of announced results.

\section{Preliminaries}

In this section we briefly remind some results of our previous paper $\left[
2\right] $ in which based on ideas of quantum theory of open systems the
formalism of probabilistic Boolean logic (PBL) was developed. To this end in
view, instead of standard Boolean variables with two values 1 and 0 that
represent true and false propositions respectively one has to consider so
called plausible propositions (PP), whose truth or falsity are determined
only with certain probability. We propose to represent such propositions by
means of $2\times 2$ diagonal matrices with positive elements the sum of
which is equal to unit. Every such representative matrix of PP may be
associated with the density matrix of certain two state quantum system .
Therefore, if it does not lead to confusion, we will often identify the
propositions in PBL with their representative density matrices (RDMs).

In explicit form the RDM of arbitrary PP $a$ looks as $\rho \left( a\right) =%
\begin{pmatrix}
p_{a} & 0 \\ 
0 & 1-p_{a}%
\end{pmatrix}%
$ where $p_{a}$ is a probability for proposition $\ a$ \ to be true(index $a$
of the proposition we will often omit later ). It is convenient also to
define the plausibility of the proposition $a$ as $P\left( a\right)
=2p_{a}-1 $. Obviously that plausibility takes its values in the interval $%
\left[ -1,1\right] $. It turns out the approach proposed allows one to
express all logical connectives between PP by universal way as certain
positive valued (PV) quantum operations under their RDMs. Referring the
reader for the details of this approach to \cite{2s} let us merely
demonstrate here how this approach works using several concrete examples.
So, if someone wants to obtain the RDM of the negation of PP $a=%
\begin{pmatrix}
p & 0 \\ 
0 & 1-p%
\end{pmatrix}%
$ that is (not $a$) she(or he) must make the next quantum operation: $\rho
\left( nota\right) =G_{not}\rho \left( a\right) G_{not}^{T}$ \ with the
following $2\times 2$ matrix $G_{not}=%
\begin{pmatrix}
0 & 1 \\ 
1 & 0%
\end{pmatrix}%
$ and $G^{T}$ is, as usually, the matrix transposed to matrix $G$. It turns
out that any two place connective in PBL $(aRb)$ may be expressed in similar
manner as well, namely:

\begin{equation}
\left( aRb\right) =G_{R}\left[ \rho \left( a\right) \otimes \rho \left(
b\right) \right] G_{R}^{T}  \label{1n5}
\end{equation}%
$\ $where $\rho \left( a\right) \otimes \rho \left( b\right) $ is standard
tensor product of RDMs of propositions $a$ and $b$ and $G_{R}$ is a $2\times
4$ matrix (we will term it below as admissible matrix for connective $R$)
which has two defining properties:1) every element of $G_{R}$ is equal $0$
or $1$ and 2) in each column of matrix $G_{R}$ \ the only element is equal
to one and all the rest are equal to zero. Let us now present (without
proof) the concrete form of matrices $G_{R}$ for basic two place
connectives, namely :$\left( a\text{ }and\text{ }b\right) $, $\left( a\text{ 
}or\text{ }b\right) $,$\left( a\Longrightarrow b\right) $ . As shown in \cite%
{2s} they have the next form:%
\begin{eqnarray}
G_{\left( a\text{ }and\text{ }b\right) } &=&%
\begin{pmatrix}
1 & 0 & 0 & 0 \\ 
0 & 1 & 1 & 1%
\end{pmatrix}%
,\  \\
G_{\left( a\text{ }or\text{ }b\right) } &=&%
\begin{pmatrix}
1 & 1 & 1 & 0 \\ 
0 & 0 & 0 & 1%
\end{pmatrix}%
,  \label{2n5} \\
G_{\left( a\Longrightarrow b\right) } &=&%
\begin{pmatrix}
1 & 0 & 1 & 1 \\ 
0 & 1 & 0 & 0%
\end{pmatrix}%
.
\end{eqnarray}
For example for the implication $\left( a\Longrightarrow b\right) $ one can
easily writes down the RDM as:%
\begin{equation}
\rho \left( a\Longrightarrow b\right) =%
\begin{pmatrix}
1-p+pq & 0 \\ 
0 & p\left( 1-q\right)%
\end{pmatrix}%
\text{.}  \label{2n5m}
\end{equation}%
In Eq. (\ref{2n5m}) we mean the next notation: $a=%
\begin{pmatrix}
p & 0 \\ 
0 & 1-p%
\end{pmatrix}%
$, and $b=%
\begin{pmatrix}
q & 0 \\ 
0 & 1-q%
\end{pmatrix}%
$. Note that this relation coincides with standard Boolean expression for
the implication in the case when probabilities $p$ and $q$ may take only two
values $0$ and $1$. Let us make now the generalization of the approach
proposed on the important case when initial PPs $a$ and $b$ are not
independent propositions or, in other words, they possess some logical
correlations. In such situation the best way to take these correlations into
account is to avail of the direct analogy between our approach and the
theory of composite correlated quantum systems. To realize this analogy one
must take as a starting point, instead of tensor production of two initial
PPs $a$ and $b$ the more general positive diagonal matrix of the form: $\rho
\left( a,b\right) =%
\begin{pmatrix}
p_{1} & 0 & 0 & 0 \\ 
0 & p_{2} & 0 & 0 \\ 
0 & 0 & p_{3} & 0 \\ 
0 & 0 & 0 & p_{4}%
\end{pmatrix}%
$ with additional normalization condition: $\sum\limits_{i=1}^{4}p_{i}=1$.
Furthermore we naturally assume that this joint matrix of correlated
propositions $\rho \left( a,b\right) $ is connected with matrices of partial
propositions that is RDM of $a$ and RDM of $b$ by the standard relations: $a=%
\begin{pmatrix}
p_{1}+p_{2} & 0 \\ 
0 & p_{3}+p_{4}%
\end{pmatrix}%
$, $b=%
\begin{pmatrix}
p_{1}+p_{3} & 0 \\ 
0 & p_{2}+p_{4}%
\end{pmatrix}%
$. Now let us suppose again that all logical connectives for correlated PP
may be obtained by the approved method that is by means of the PV quantum
operations. Here we restrict ourselves only with the case of implication
between correlated propositions $a$ and $b$. Let us introduce the following
notation $p=p_{1}+p_{2}$ , $q=p_{1}+p_{3}$ and $C=p_{1}p_{4}-p_{2}p_{3}$

It is easy to see that ,using this notation, the matrix $\rho \left(
a,b\right) $ can be rewritten in the form: $\rho \left( a,b\right) =%
\begin{pmatrix}
pq+C & 0 & 0 & 0 \\ 
0 & p\left( 1-q\right) -C & 0 & 0 \\ 
0 & 0 & q\left( 1-p\right) -C & 0 \\ 
0 & 0 & 0 & \left( 1-p\right) \left( 1-q\right) +C%
\end{pmatrix}%
$. Now by applying the quantum operation $G_{\left( a\Longrightarrow
b\right) }=%
\begin{pmatrix}
1 & 0 & 1 & 1 \\ 
0 & 1 & 0 & 0%
\end{pmatrix}%
$ to the $\rho \left( a,b\right) $ one may obtain the required result for
RDM of implication, namely:%
\begin{equation}
\rho \left( a,b\right) =%
\begin{pmatrix}
1-p+pq+C & 0 \\ 
0 & p-pq-C%
\end{pmatrix}%
\text{.}  \label{4n5}
\end{equation}%
Note that the variable $C$ is just a measure of logical correlations between
propositions $a$ and $b$. In this connection we will termed it as the
context variable.

Let us consider now the simple example with three concrete initial RDMs of
composite correlated propositions, namely :

1) $\rho _{1}\left( a,b\right) =%
\begin{pmatrix}
\frac{1}{2} & 0 & 0 & 0 \\ 
0 & 0 & 0 & 0 \\ 
0 & 0 & 0 & 0 \\ 
0 & 0 & 0 & \frac{1}{2}%
\end{pmatrix}%
$,

2) $\rho _{2}\left( a,b\right) =%
\begin{pmatrix}
0 & 0 & 0 & 0 \\ 
0 & \frac{1}{2} & 0 & 0 \\ 
0 & 0 & \frac{1}{2} & 0 \\ 
0 & 0 & 0 & 0%
\end{pmatrix}%
$ and

3) $\rho _{3}=%
\begin{pmatrix}
\frac{1}{4} & 0 & 0 & 0 \\ 
0 & \frac{1}{4} & 0 & 0 \\ 
0 & 0 & \frac{1}{4} & 0 \\ 
0 & 0 & 0 & \frac{1}{4}%
\end{pmatrix}%
$.

It is easy to see that in all cases the RDMs for partial propositions $a$
and $b$ coincide and represent the same ambiguous proposition: $\rho
_{i}\left( a\right) =\rho _{i}\left( b\right) =%
\begin{pmatrix}
\frac{1}{2} & 0 \\ 
0 & \frac{1}{2}%
\end{pmatrix}%
\left( i=1,2,3\right) $. However the expressions for implication in these
situations differ significantly. Really using the general rule Eq. (\ref{1n5}%
) stated above and its special case in the form of Eq. (\ref{4n5}) one may
easily obtain: $\left( a\Longrightarrow b\right) _{1}=%
\begin{pmatrix}
1 & 0 \\ 
0 & 0%
\end{pmatrix}%
$, that is the true proposition, while $\left( a\Longrightarrow b\right)
_{2}=%
\begin{pmatrix}
\frac{1}{2} & 0 \\ 
0 & \frac{1}{2}%
\end{pmatrix}%
$ is the same ambiguous proposition and $\left( a\Longrightarrow b\right)
_{3}=%
\begin{pmatrix}
\frac{3}{4} & 0 \\ 
0 & \frac{1}{4}%
\end{pmatrix}%
$ .Note that only the last case with composite matrix $\rho _{3}$
corresponds to the case of the independent partial propositions. This simple
but instructive example clearly demonstrates the considerable influence of
logical correlations on the results of inference process. In this point we
have finished our brief review of PBL. It should be noted in conclusion that
in the case of PBL considered in this Section were taken into account only
propositions that may be specified by the single defining characteristic ,
namely plausibility. Now we pass to the study of Non-Boolean (NB) logics
that would operate with the propositions of more general kind .These
propositions can be determined already by two defining features which
reflect dual nature of the objects (or subjects) connecting with
corresponding propositions . We believe that similar logics may be more
appropriate for the description (at least in simplified form) of essential
peculiarities of such common logics as the logic of primitive (mythological)
thinking and the logic of everyday human reasoning.

\section{Non-Boolean logic of the first kind as the simplified description
of primitive (mythological) thinking}

In the beginning of this Section we would like to discuss the following
primary question namely: to what degree the approach stated above may be
extended on the propositions of more general form .To this end in view let
us consider instead of exclusively diagonal RDM representing a certain PP
the more general class of $2\times 2$ non-diagonal positive valued matrices
with unit trace and examine them as certain RDMs of generalized propositions
(GP). Let us suppose that an arbitrary proposition of this class $a$ can be
written in the next form $\rho \left( a\right) =%
\begin{pmatrix}
p & z \\ 
z^{\ast } & 1-p%
\end{pmatrix}%
$ ,where $z$ is some complex number. Let us define also the negation of such
proposition, that is $\left( not\text{ }a\right) $ by natural way as $\rho
\left( not\text{ }a\right) =%
\begin{pmatrix}
1-p & z^{\ast } \\ 
z & p%
\end{pmatrix}%
$. Furthermore let us require that the RDMs of the opposite propositions $a$
and $not$ $a$ commute with each other. The sense of this restriction (having
in mind the obvious analogy with quantum theory) is quite clear and does not
need in additional comments. It is easy to see that this restriction may be
reduced to the simple relation $z=i\alpha $ or, in other words, non-diagonal
elements of the RDM in the approach proposed have to be pure imaginary. Thus
the class of generalized RDMs which we will consider in the present Section
contains all $2\times 2$ positive valued matrices of the form:%
\begin{equation}
\rho \left( a\right) =%
\begin{pmatrix}
p & i\alpha  \\ 
-i\alpha  & 1-p%
\end{pmatrix}%
\text{.}  \label{5n5}
\end{equation}%
As to formal interpretation of these GPs we assume as before that their
diagonal elements determine the plausibility of corresponding GP according
to the above mentioned rule: $P\left( a\right) =2p-1$ whereas their
non-diagonal elements define its believability, that is the readiness of a
subject to accept the corresponding proposition as valid by virtue of some
irrational and unconscious reasons or, in other words. on the strength of
certain beliefs. Thus the similar GPs combine together the objective and
subjective reasons which stimulate a person to accept the concrete
proposition as valid. It is naturally to define the believability of
concrete GP $a$ with RDM Eq. (\ref{7n5}) as $B\left( a\right) =-2\alpha $.
The justification of such definition became clear if we write down this GP
by standard way using the Bloch sphere representation, namely:%
\begin{equation}
\rho \left( a\right) =%
\begin{pmatrix}
\frac{1+P_{z}}{2} & \frac{P_{x}-iP_{y}}{2} \\ 
\frac{P_{x}+iP_{y}}{2} & \frac{1-P_{z}}{2}%
\end{pmatrix}%
\text{.}  \label{6n5m}
\end{equation}

Note that for GPs which we consider here, the component $P_{x\text{ }}$of
the Bloch vector $\overrightarrow{P}$ is always equal to zero. Now if one
compares the representation Eq. (\ref{6n5m}) with expression Eq. (\ref{5n5})
she(he) can see that the plausibility and believability of the same GP being
expressed in terms of the Bloch vector components look similarly. Furthemore
it should be noted that relation $P_{z}^{2}+P_{y}^{2}$ $\leqslant 1$ is
obviously holds for any GP and this fact distinctly reflects the competition
existing between two features of the same proposition. It should be noted
also that our assumption is in a full agreement with modern psychological
dual system theory \cite{3s} that asserts the presence in brain two
complementary cognitive systems: (deductive and heuristic) , which both are
responsible for human reasoning, although often compete with each other. Now
in order to formulate the consistent NB logic with GPs stated above we have
to define all possible connectives related to them. It turns out however
that the direct extension of the admissible quantum operation method in
order to obtain required two place logical operations for GPs is impossible
( because the automatic satisfaction of normalization condition, that was
guaranteed for PPs in PBL, is not always holds now).However there are at
least two possible ways to get round this formal obstacle. In this Section
we consider the first of them. To this end we retain as admissible
operations only two concrete $2\times 4$ matrices,namely: $G_{\Delta }=%
\begin{pmatrix}
0 & 1 & 1 & 0 \\ 
1 & 0 & 0 & 1%
\end{pmatrix}%
$ and $G_{\overline{\Delta }}=%
\begin{pmatrix}
1 & 0 & 0 & 1 \\ 
0 & 1 & 1 & 0%
\end{pmatrix}%
$. It is easy to see directly that these operations transform the tensor
product of two propositions $a\otimes b$ to the RDM that belongs to the
required class of GP . Really, let $a=%
\begin{pmatrix}
p & i\alpha  \\ 
-i\alpha  & 1-p%
\end{pmatrix}%
$ and $b=%
\begin{pmatrix}
q & i\beta  \\ 
-i\beta  & 1-q%
\end{pmatrix}%
$ are two GP, then it is easy to obtain that:%
\begin{equation}
G_{\Delta }\left( a\otimes b\right) G_{\Delta }^{T}=%
\begin{pmatrix}
A & i\gamma  \\ 
-i\gamma  & 1-A%
\end{pmatrix}%
\equiv \left( a\Delta b\right) \text{.}  \label{7n5}
\end{equation}%
where $A=p+q-2pq+2\alpha \beta $ and $\gamma =\alpha \left( 1-2q\right)
+\beta \left( 1-2p\right) $. In the same way one can find that%
\begin{eqnarray}
G_{\overline{\Delta }}\left( a\otimes b\right) G_{\overline{\Delta }}^{T} &=&%
\begin{pmatrix}
1-A & -i\gamma  \\ 
i\gamma  & A%
\end{pmatrix}%
=  \label{8n5} \\
&=&not(a\Delta b)\equiv (a\overline{\Delta }b)\text{.}  \notag
\end{eqnarray}

In the case when $x$-component of the Bloch vector is identically equal to
zero it is convenient to introduce the complex Bloch vector as follows : $%
P=P_{z}-iP_{y}.$Comparing this definition with the above mentioned Bloch
sphere representation of density matrix one can state the simple relation
connecting the complex vector $R$ of the proposition $\left( a\Delta
b\right) $with corresponding vectors $P$ and $Q$ of the GP $a$ and $b$
respectively, namely:%
\begin{equation}
R=-PQ\text{.}  \label{11n5m}
\end{equation}
So, we found that the NB logic of the first kind aside from negation
contains in addition only the single pair of two place connectives, namely $%
\Delta $ and $\overline{\Delta }$. At first sight such logic is much poorer
that standard Boolean logic and therefore one may call it as 'atomic logic'\
or prime logic. Nevertheless it should be noted that, having in hands only
these two connectives and starting from the set of GPs, one can easily
obtain any PP belonging to PBL with prescribed diagonal elements. After
making that, one may operate with them already according to the rules of
standard Boolean logic. Let us demonstrate the validity of this statement
.To this end let us take two opposite GP of the special form: $a=%
\begin{pmatrix}
\frac{1}{2} & i\alpha \\ 
-i\alpha & \frac{1}{2}%
\end{pmatrix}%
$ and $\left( not\text{ }a\right) =%
\begin{pmatrix}
\frac{1}{2} & -i\alpha \\ 
i\alpha & \frac{1}{2}%
\end{pmatrix}%
$. If one applies the operation $\Delta $ to them the obtained result looks
as:%
\begin{equation}
\left( a\Delta \text{ }not\text{ }a\right) =%
\begin{pmatrix}
\frac{1}{2}-2\alpha ^{2} & 0 \\ 
0 & \frac{1}{2}+2\alpha ^{2}%
\end{pmatrix}%
\text{.}  \label{9n5}
\end{equation}

Thus we see that any proposition $a$ of PBL with $p_{a}\leqslant \frac{1}{2}$
may be expressed by similar way. As for propositions of PBL with $p_{a}\geq 
\frac{1}{2}$ \ they can be expressed by means of operation $\overline{\Delta 
}$ in the same manner. Thus the PBL plays now the role of secondary or
'molecular' logic\ with respect to prime or 'atomic'\ NB logic of the first
kind. Note in addition that the GP of the form : $a=%
\begin{pmatrix}
\frac{1}{2} & i\alpha \\ 
-i\alpha & \frac{1}{2}%
\end{pmatrix}%
$ whose plausibility is equal to zero can be considered as the simplest
example of so-called intuitive judgments. In this place we refer the reader
to the inspiring book \cite{4s} in which distinction between discoursive and
intuitive judgments have been deeply analyzed. Now let us pass to the most
intriguing question, relating to the NB logic of the first kind stated
above, namely: whether the logic proposed above has any actual applications
in everyday life and (or) in science? We believe that the answer is yes and
are going to argue that logic of this kind could be appropriate for example
as simplified description of primitive ( mythological) thinking. The author
is aware that without being an expert in the area of anthropology or
mythology he is unable to disclose this topic in proper depth and moreover
to prove the above assertion exactly. Nevertheless let me adduce several
arguments in favor of the hypothesis proposed above. In the beginning let
usremind some basic facts relating to the primitive thinking (PT), or the
thinking of savages ( all these and the other concrete facts the reader can
derive in detail for example from classical book by C. Levi-Strauss) \cite%
{5s}.The major fact on which we base is that there are a set of elementary
and at the same time fundamental units of PT, so- called binary oppositions,
such as : top-bottom, right-left, birth-death, male-female and so on.
Another important regularity determines the basic principle governing the
primitive thinking, that is, its goals and mental tools which are used to
achieve these goals . It turns out that PT while it operates with binary
oppositions does not seek to avoid any contradictions, that is typical for
the "normal\textquotedblright\ logic, but rather tends to find and then to
reconstruct all kinds of possible intermediate links existing between two
antagonistic terms in concrete binary opposition. To this end in view there
are two special personages in myth, namely: the Culture Hero and the
Trixter, (for example in Greek mythology Hermes plays the role of such
Trixter) which realize this important function. Furthermore certain totem
animals for example the Raven and the Coyote in Indian myths sometimes play
the part of Culture hero as well. The basic tool using in PT that allows one
to achieve the desired goal is so-called 'mediation'\ --the special mental
or workable operation which is able to reduce the distinction between two
antagonistic terms in binary oppositions. For example the Raven and the
Coyote as the animals- scavengers symbolically realize the mediation
function between the world of the living beings and the world of the dead .
Now let us demonstrate how in the NB logic of the first kind described above
its two basic operations may be considered as certain analogues of the
mediation operation in myth. To this end, in order to define the degree of
difference between two distinct GPs, we will use well-known definition of
the distance between two mixed quantum states $\rho _{1\text{ }}$and $\rho
_{2},$ namely $D\left( \rho _{1},\rho _{2}\right) =\frac{1}{2}Sp\left\vert
\rho _{1}-\rho _{2}\right\vert $ \cite{6s}. It is easy to see that in the
case of two dimensional Hilbert space and in the Bloch sphere representation
the above expression takes the simple form: $D\left( \rho _{1},\rho
_{2}\right) =\frac{1}{2}\left\vert \overrightarrow{P}_{1}-\overrightarrow{P}%
_{2}\right\vert $ where $\overrightarrow{P}_{1\text{ \ }}$and $%
\overrightarrow{P}_{2}$ are corresponding Bloch vectors of the states.

It is obvious that maximum distance is realized between opposite
propositions or , (using the PT language) between antagonistic members of
binary opposition. Now if one applies the basic two place operation $\Delta $
to the pair of opposite propositions $a$ with complex vector $P$ and $\left(
nota\right) $ with complex vector $-P$ she(he) obtains the new proposition
with the complex vector $R=P^{2}$. It is easy to see that the distance
between the opposite propositions with complex vectors $P^{2}$ and $-P^{2}$
is less that the distance between original pair of opposite propositions.
This fact confirms that logical connective $\Delta $ possesses the
characteristic property of operation 'mediation'\ in PT. Now we are going to
adduce another forcible argument in favor of the doubtless connection
between NB logic of the first kind considered above and the logic of PT. Let
us remind that according to the known ethnologist Levy-Bruhl \cite{7s} (who
was the first European researcher of primitive and magical thinking PT is
governed to a great extent by the so-called law of participation that claims
the universal links existing between various things and events in the
world.In particular just this law allows the shaman, who fell into trans, to
perform a wide variety of magical acts and transformations with surrounding
objects or people using their mutual contiguity and (or) similarity. It is
interesting to note that this feature of primitive thinking afterwards was
adopted by art and literature (especially poetry) in the form of extensive
use of such specific tropes as metaphor and metonymy. In the NB logic of the
first kind proposed above this feature of PT may be simulated by additional
logic operations,namely by logic rotations, that have no analogues in the
standard Boolean logic. Really, it is possible to define one-parameter
continuous group of logic rotations (in the plane $P_{z}-P_{y}$) according
to the usual mathematical rule, namely: let GP $a$ is represented by the
complex vector $P$ then GP $\widetilde{a}$ obtained by rotation of $a$ at an
angle $\Phi $ has complex vector $\widetilde{P}$ with components $\widetilde{%
P_{y}}$ and $\widetilde{P_{z}}$ so that $\widetilde{P_{y}}=P_{y}\cos \Phi
+P_{z}\sin \Phi $ and $\widetilde{P_{z}}=P_{z}\cos \Phi -P_{y}\sin \Phi $
.Clearly that in this groupof logical transformations the negation of the
proposition $a$ coincides with its rotation at angle $\pi $ and,
furthermore, if one rotates GP $a$ at angle $\Phi _{1}$ and the other GP $b$
at angle $\Phi _{2}$ then the GP $\left( a\Delta b\right) $ would be rotated
at angle $(\Phi _{1}+\Phi _{2})$.Thus this continuous group of logical \
rotations let one to transform any proposition into the arbitrary other
(with the same modulus certainly). In this point we are obliged to finish
our brief discussion of the relationship existing between primitive thinking
and the NB logic of the first kind although it is clear that this complex
and intriguing topic deserves much more detail study.

\section{Non-Boolean logic of the second kind as the toy model of everyday
human reasoning}

As we see in the previous Section in the case of Non- Boolean logic of the
first kind it is possible to define aside from negation only the single pair
of two- place logical operations $\Delta $ and $\overline{\Delta }$. The
application of the rest admissible quantum operations to the tensor product
of two input GP results in the disturbance of normalization condition for
output RDM. This obstacle may be formally removed if one assume the
existence of certain primordial correlations between plausible and
believable components of the GPs. Let us explain more detail what we have in
mind. Suppose that instead of tensor product of the two input GP $a\otimes b$
(that means their logical independence) we take as input more general
composite proposition $\rho \left( a,b\right) $ , the RDM of which has the
next form:%
\begin{equation}
\rho \left( a,b\right) =%
\begin{pmatrix}
p_{1} & i\alpha & i\beta & 0 \\ 
-i\alpha & p_{2} & 0 & i\gamma \\ 
-i\beta & 0 & p_{3} & i\delta \\ 
0 & -i\gamma & -i\delta & p_{4}%
\end{pmatrix}%
\text{.}  \label{10n5}
\end{equation}%
Using the obvious analogy with theory of composite quantum systems one can
determine now the partial RDM of propositions $a$ and $b$ as $a=%
\begin{pmatrix}
p_{1}+p_{2} & i\left( \beta +\gamma \right) \\ 
-i\left( \beta +\gamma \right) & p_{3}+p_{4}%
\end{pmatrix}%
$ and $b=%
\begin{pmatrix}
p_{1}+p_{3} & i\left( \alpha +\delta \right) \\ 
-i\left( \alpha +\delta \right) & p_{2}+p_{4}%
\end{pmatrix}%
$ respectively. It is clear that input composite proposition $\rho \left(
a,b\right) $ Eq. (\ref{10n5}) corresponds to the case of two correlated
partial GP $a$ and $b$ and the degree and nature of their correlation is
uniquely determined by elements of composite matrix $\rho \left( a,b\right) $%
. It should be emphasized that construction of the input composite
proposition $\rho \left( a,b\right) $ in the form Eq. (\ref{10n5}) has the
essential advantage over simple tensor product from formal point of view
since it allows one to define all 16 two place logical connectives
consistently by approved method of PV quantum operations. Significantly that
all admissible matrices $G_{R}$ in this NB logic of the second kind have the
same form as in the case of PBL (see Eq. (\ref{1n5}) ) and corresponding two
place connectives can be obtained according to the similar rule:%
\begin{equation}
\left( aRb\right) =G_{R}\rho \left( a,b\right) G_{R}^{T}\text{.}
\label{11n5}
\end{equation}%
Let us present here (without details) the expressions for the two place
connectives $\left( and\right) $, $\left( or\right) $ in the NB logic of the
second kind, obtained by means of Eq. (\ref{11n5}):%
\begin{eqnarray}
\left( a\text{ }and\text{ }b\right) &=&%
\begin{pmatrix}
p_{1} & i\left( \alpha +\beta \right) \\ 
-i\left( \alpha +\beta \right) & p_{2}+p_{3}+p_{4}%
\end{pmatrix}%
,  \label{12n5} \\
\left( a\text{ }or\text{ }b\right) &=&%
\begin{pmatrix}
p_{1}+p_{2}+p_{3} & i\left( \gamma +\delta \right) \\ 
-i\left( \gamma +\delta \right) & p_{4}%
\end{pmatrix}%
\end{eqnarray}%
Now as in the previous section, the important question arises about the
possibility of concrete realizations of the NB logic of the second kind.

In this connection we are going to bring several arguments in favor of
hypothesis that above version of NB logic can be considered as simplified
model of everyday human reasoning. To this end in view let us remind some
basic facts relating to the psychological dual system theory of human
cognition(see for example \cite{3s} for the comprehensive exposition). As
numerous psychological experiments and observations had unambiguously
demonstrated there are two distinct types of mental processes that are
responsible for human reasoning: 1) cognitive processes of the first type
that are rapid, automatic and intuitive and 2) the processes of the second
type that are slow, sequent and rational. Cognitive system which is
responsible for the processes of the first type is termed as heuristic (or
intuitive) while the second one is termed as deductive (or analytical).
Clearly these two systems may sometimes come into conflict and compete with
each other. Therefore human reasoning is often biased and subjected to
various fallacies. In particular the reasoning person often overestimates
the cogence of the arguments that lead to the believable and expected
conclusions and underestimates those arguments which lead to the conclusions
contradictory to her (his) beliefs. This widespread psychological phenomenon
is known as belief-bias effect. Now the natural question arises: is it
possible to describe this human hybrid intuitive-deductive thinking in the
framework of any consistent logical theory? In this Section we propose brief
outline of possible theory based on the NB logic of the second kind that was
exposed above. Note that we present here only the little fragment of this
theory which let one to understand belief-bias effect in the framework of
closed logical point of view. To this end let us examine the expression for
implication connecting two correlated GP. According to the general rule Eq. (%
\ref{11n5}) the RDM of the implication can be represented in the form:%
\begin{equation}
\left( a\Longrightarrow b\right) =G_{imp}\rho \left( a,b\right) G_{imp}^{T}=%
\begin{pmatrix}
p_{1}+p_{3}+p_{4} & i\left( \alpha -\gamma \right) \\ 
-i\left( \alpha -\gamma \right) & p_{2}%
\end{pmatrix}%
\text{.}  \label{13n5}
\end{equation}%
where the matrix $G_{imp}=%
\begin{pmatrix}
1 & 0 & 1 & 1 \\ 
0 & 1 & 0 & 0%
\end{pmatrix}%
$ .

We see that the believability of the GP $\left( a\Longrightarrow b\right) $
as it follows from Eq. (\ref{13n5}) is equal to $2\left( \gamma -\alpha
\right) $, and it does not connect by simple way with corresponding
believabilities of partial propositions $a$ and $b$. Let us now impose
another additional restrictionon on the phases $\alpha ,\beta ,\gamma
,\delta $ of the matrix $\rho \left( a,b\right) $ which looks enough
naturally, namely, we require that in simple version of NB logic of the
second kind the known De Morgan dual formulas for all logical propositions,
namely: $not\left( a\text{ }and\text{ }b\right) =\left( not\text{ }a\right) $
$or\left( not\text{ }b\right) $ and $not\left( a\text{ }or\text{ }b\right)
=\left( not\text{ }a\right) $ $and$ $\left( not\text{ }b\right) $ hold. This
is just the case what we have in mind when we are speaking about the toy
model of everyday human reasoning. Really, it is easy to see that the
restriction above immediately leads to the next condition imposed on the
phases of input matrix $\rho \left( a,b\right) $ of Eq. (\ref{10n5}),
namely: $\alpha +\beta =\gamma +\delta =const$. Since this constant must be
identical for all propositions in the toy model of NB logic of the second
kind we can without loss of generality set it equal to zero and then it
turns out that all non-diagonal elements in implication ( and in other
connectives as well) can be expressed by means of the non-diagonal elements
of partial propositions. Comparing the believability of the implication $%
\left( a\Longrightarrow b\right) $ with the believability of it consequence $%
b$ that is equal to $-2\left( \alpha +\delta \right) =2\left( \gamma -\alpha
\right) $we immediately come to the desired conclusion , namely, in the toy
version of NB logic of the second kind the values of believabilities of the
propositions $\left( a\Longrightarrow b\right) $ and $b$ completely coincide
(and both are equal to $2\left( \gamma -\alpha \right) )$.Strictly speaking
just this conclusion gives us the good reason to believe that NB logic of
the second kind has essential common features with everyday human logic in
which (as was shown in numerous experiments) the believability of
conclusions stimulates the subjects more positively evaluate the correctness
of the process of their inference and to pay less attention to their logical
rigor (belief-bias effect). It is necessary to mention also that as a matter
of fact the whole subjective evaluation of the validity of the given
proposition $a$ simultaneously should take into account both the
plausibility of this proposition and its believability as well. Therefore
the appropriate expression for the integral validity $V$ of the GP $a$
should be actually look as follows:%
\begin{equation}
V\left( a\right) =\alpha P\left( a\right) +\left( 1-\alpha \right) B\left(
a\right) \text{.}  \label{14n5}
\end{equation}%
(where the coefficient $\alpha $ reflects the mental nature of reasoning
person.It remains to point out$\ \ \ $that by application of the results of
the approach proposed in this Section one can evaluate quantitatively both
believability and plausibility of various GPs in complex logical situations
relating to everyday human reasoning. However this amusing task is already
beyond the scope of the present paper.

In the conclusion we should like to emphasize only that all guesses and
conjectures about proposed interpretations of \ two NB logics expressed in
this paper must be taken with a grain of salt, that is as plausible but
preliminary hypotheses . Obviously, the rigorous confirmation of all results
obtained here will require more painstaking extra work in the future.

\end{document}